\documentclass[twocolumn]{aastex631}
\usepackage{CJK}

\shorttitle{Anisotropic Absorption from A Single Antenna}
\shortauthors{Ao et al.}
\graphicspath{{./}{figures/}}
\usepackage{amsmath}
\usepackage{amssymb}
\usepackage{graphicx}
\usepackage{multirow}
\usepackage{subfigure}

\begin{document}
\begin{CJK*}{UTF8}{gbsn} 
\title{Reconstructing the Anisotropic Ultra-long Wavelength Spectra using a Single Antenna on Lunar-orbit}

\correspondingauthor{Yidong Xu; Furen Deng}
\email{xuyd@nao.cas.cn; frdeng@nao.cas.cn}

\author[0009-0007-2297-3718]{Qige Ao (奥琦格)}
\affiliation{National Astronomical Observatories, Chinese Academy of Sciences, 20A Datun Road, Chaoyang District, Beijing 100101, China}
\affiliation{Shanghai Astronomical Observatory, Chinese Academy of Sciences, 80 Nandan Road, Xuhui District, 200030, China}
\affiliation{School of Astronomy and Space Science, University of Chinese Academy of Sciences, No.1 Yanqihu East Rd, Huairou District, Beijing 101408, China}

\author[0000-0001-8075-0909]{Furen Deng (邓辅仁)}
\affiliation{National Astronomical Observatories, Chinese Academy of Sciences, 20A Datun Road, Chaoyang District, Beijing 100101, China}
\affiliation{School of Astronomy and Space Science, University of Chinese Academy of Sciences, No.1 Yanqihu East Rd, Huairou District, Beijing 101408, China}
\affiliation{State Key Laboratory of Radio Astronomy and Technology, Beijing 100101, People's Republic of China}

\author[0000-0003-3224-4125]{Yidong Xu (徐怡冬)}
\affiliation{National Astronomical Observatories, Chinese Academy of Sciences, 20A Datun Road, Chaoyang District, Beijing 100101, China}
\affiliation{State Key Laboratory of Radio Astronomy and Technology, Beijing 100101, People's Republic of China}

\author[0000-0002-7829-1181]{Bin Yue (岳斌)}
\affiliation{National Astronomical Observatories, Chinese Academy of Sciences, 20A Datun Road, Chaoyang District, Beijing 100101, China}
\affiliation{State Key Laboratory of Radio Astronomy and Technology, Beijing 100101, People's Republic of China}

\author[0000-0001-8534-837X]{Huanyuan Shan (陕欢源)}
\affiliation{Shanghai Astronomical Observatory, Chinese Academy of Sciences, 80 Nandan Road, Xuhui District, 200030, China}
\affiliation{School of Astronomy and Space Science, University of Chinese Academy of Sciences, No.1 Yanqihu East Rd, Huairou District, Beijing 101408, China}
\affiliation{State Key Laboratory of Radio Astronomy and Technology, Beijing 100101, People's Republic of China}

\author[0000-0001-6475-8863]{Xuelei Chen (陈学雷)}
\affiliation{National Astronomical Observatories, Chinese Academy of Sciences, 20A Datun Road, Chaoyang District, Beijing 100101, China}
\affiliation{School of Astronomy and Space Science, University of Chinese Academy of Sciences, No.1 Yanqihu East Rd, Huairou District, Beijing 101408, China}
\affiliation{State Key Laboratory of Radio Astronomy and Technology, Beijing 100101, People's Republic of China}

\shorttitle{Anisotropic Absorption from a Single Antenna}
\shortauthors{Ao et al.}

\begin{abstract}

The ultra-long wavelength sky ($\nu\lesssim 30$ MHz) is still largely unexplored, as the electromagnetic wave is heavily absorbed and distorted by the ionosphere on Earth. The far-side of the Moon, either in lunar-orbit or on lunar-surface, is the ideal site for observations in this band, and the upcoming Moon-based interferometers will obtain multi-frequency high-resolution sky maps. Making use of the lunar occultation of the sky and the anisotropy of antenna primary beam response, we propose a novel method to reconstruct the ultra-long wavelength spectral shape in multiple directions in the sky using only one antenna on lunar orbit. We apply the method to one antenna on one of the nine daughter satellites of the proposed Discovering the Sky at Longest wavelength (DSL) project. Using simulated observation data between 1 - 30 MHz from one dipole antenna, we find that the spectra for different regions on the sky can be reconstructed very well and the free-free absorption feature in each region can be derived from the reconstructed spectra. This work demonstrates the feasibility of reconstructing the unbiased anisotropic spectra using
very limited instrumentation on a lunar-orbit, with mature technologies already in place. It extends the application of such kind of satellite in revealing the distribution of free electrons in the Galactic interstellar medium from the distribution of absorption features in the ultra-long wavelength sky.

\end{abstract}


\section{Introduction} \label{sec:intro}

The ultra-long wavelength band (wavelength $\lambda \gtrsim$ 10 m or frequency $\nu\lesssim$ 30 MHz) is the remaining electromagnetic window that is still largely unexplored, particularly below $\lesssim$ 10 MHz (e.g., \citealt{brown_galactic_1973,Cane_1979,Yates_1967,novaco_nonthermal_1978,Peterson_2002}). Some proposed projects, for example FARSIDE \citep{FARSIDE}, DAPPER \citep{DAPPER}, LuSEE-Lite/LuSEE-Night \citep{Lusee}, FarView \citep{FarView2024AdSpR}, etc., 
aim to carry out observations in the ultra-long wavelength band in order to uncover new celestial objects and physical mechanisms in this new window, and to understand the foregrounds for 21-cm cosmology from the Dark Ages.
Among these, the Discovering the Sky at Longest wavelength (DSL) is a proposed upcoming lunar-orbit satellite array project. It is composed of one mother satellite, eight low-frequency daughter satellites that form a linear interferometer array to take images of the Universe between 0.1 MHz and 30 MHz, and another high-frequency satellite dedicated to precisely measure the global spectrum from 30 - 120 MHz \citep{Chen_2019,chen_discovering_2021}.   
For the eight low-frequency satellites, each of them carries three pairs of orthogonal monopole antennas on the two sides of the satellite. Each pair of antennas acts as a short dipole that connects to an independent channel of the receiver, so they can work either as an interferometer or independently.

Early observations revealed a downturn in 
the global spectrum of radio background around $\sim$ 3--5 MHz, which has been attributed to free-free absorption by free electrons in the interstellar medium (ISM) of our Milky Way (e.g., \citealt{ellis_cosmic_1966,alexander_rocket_1965,smith_cosmic_1965,alexander_spectrum_1969,brown_galactic_1973,novaco_nonthermal_1978}).
This is also found in the frequency spectrum of the radio sky anisotropy \citep{Page2022AA} and can put constraints on sky models that involve free-free absorption \citep{Bassett2023ApJ}. \citet{cong_ultralong-wavelength_2021} developed a radio sky model, the Ultra-Long wavelength Sky model with Absorption ({\tt ULSA}), which involves this absorption based on Milky Way electron model \citep{Ne2001_1,Ne2001_2}, so it is valid down to $\lesssim 10$ MHz. 
{\tt ULSA} predicts that, in the ultra-long wavelength band, the sky morphology is rather different from higher frequencies. The Galactic plane is darker than high Galactic latitude regions, and the electron structures may leave shadows and bright spots on the sky. This absorption mechanism provides an opportunity to reconstruct the 3-D free-electron distribution from ultra-long wavelength observations \citep{Cong2022ApJ}.

Although the DSL low-frequency daughter satellites are designed to make images of the sky from cross-correlation measurements between them, each of them also individually records the global spectrum (auto-correlation) from 0.1 to 30 MHz with high frequency resolution. The observed global spectrum varies with the pointing direction of the antenna, not only due to the non-isotropic primary beam response, but more importantly, the Moon blocks different parts of the sky as the satellite moves in orbit. 
As the satellite orbits the Moon, 
the antenna receives spectra from different sky regions, encoding spatial information into the temporal variations of the observed data. 
This implies that even in the single antenna mode, a lunar-orbit satellite has the potential to resolve the anisotropic spectra in the radio foreground. 
This spatial information can be similarly obtained by a single-antenna experiment on the lunar surface making use of lunar rotation. Employing
a Wiener filter map-making technique, \citet{LuSEE-Night2025mapmaking} have demonstrated that a low-resolution sky map can be reliably reconstructed across 5 -- 50 MHz using drift scan observations with four monopole antennas equipped on the LuSEE-Night experiment. When measurements across multiple frequencies are combined,
the anisotropic spectral shape could potentially reveal the anisotropic absorption features caused by free-free absorption.

Existing works of retrieving a low-resolution sky map as a function of frequency using a single antenna have been focused on extracting the 21-cm global spectrum from the Cosmic Dawn. In order to extract the cosmic 21-cm spectral feature, \citet{Anstey2021} employed a physically motivated formalism to model and parameterize the radio foreground by dividing the full sky into distinct regions, and developed a general pipeline to separate the 21-cm signal and reconstruct the anisotropic foreground spectrum at the same time. Based on this work, \citet{Hibbard2023} evaluated the accuracy of the fit of seven foreground models using simulated measurement data. All of these works focused on the 21-cm signal extraction in the 50 -- 200 MHz frequency band, for which the free-free absorption effect is still negligible.

For lower frequencies at $\lesssim 30$ MHz, the sky is poorly constrained, while accurate characterization of the sky itself is critical to 21-cm cosmology from the Dark Ages. The main purpose of this work is to reconstruct the foreground sky itself at frequencies lower than 30 MHz, to understand the foregrounds in this poorly constrained band, and to try to reveal the anisotropic absorption feature using very limited instrumentation in the near future.

For the purpose of extracting the 21-cm signal, previous works require the observation data to be taken with restricted Earth rotation. In particular, \citet{Anstey2023} found variations in the fitted spectral index parameters when the sky has rotated by a significant angle, which could potentially result in bias in the fitted foreground parameters as well as in the recovered 21-cm signal. Therefore, in those approaches, one needs to {\it restrict the sky variation} in order to obtain unbiased results. In contrast, here we {\it take advantages of the variation} of the sky when the Moon blocks different parts of the sky, to extract the spatial information from the time-ordered data (TOD) as the satellite orbits the Moon. 

In order to reconstruct the anisotropic spectrum caused by different free-free absorption features in different directions, we divide the sky into different regions according to the expected free-free absorption properties instead of the spectral indices as in \citet{Anstey2021}. This strategy would result in greater inhomogeneity in the spectral index of the sky in each region, and therefore a potentially more prominent bias in the fitting results. In order to minimize bias in the fitted sky parameters, one has to take into account the full covariance of errors because the TODs are inherently correlated \citep{LiHaoran2025}. In the present work, rather than evaluating the covariance in a simulation-based way as in \citet{LiHaoran2025}, we develop an analytic formalism to model different sources of errors, i. e. the thermal noise, the error due to data averaging, the error due to coarse discretization when using an average spectrum to represent the spectra of an entire region, and the spectrum model error. Using mock data simulated for a whole precession period of 1.3 yr and by accounting for the full covariance of various errors, we will be able to obtain unbiased fitting results for the low-frequency foreground sky itself.

In this paper, we investigate the feasibility of reconstructing parameterized spectra for different directions in the sky, from mock observation data from a single antenna on one of the DSL low-frequency satellites. If applied to real data, the free-free absorption to ultra-long wavelength radiation from different sky directions can be derived from the reconstructed anisotropic spectra, and the spatial distribution of the ISM free electrons can be learned from the single antenna observations in a lunar orbit. The layout of this paper is as follows. We first introduce our method to reconstruct the spectra of different sky regions in Section \ref{sec:methods}, and then in Section \ref{sec:results} we present our results of the estimated parameter values and uncertainties. In Section \ref{sec:CC}, we summarize and discuss our results.

\section{Methods} 
\label{sec:methods}

In this section, we first generate mock TOD of sky temperature from the {\tt ULSA} sky model \citep{cong_ultralong-wavelength_2021}, considering the satellite's orbital motion and natural precession, the varying Moon blockage, antenna pointing, and the beam response. By design, the DSL low-frequency satellites can measure the electromagnetic radiation down to 0.1 MHz, however, the spectral shape may become complicated below $\sim$ 1 MHz, because at such low frequencies, the radiation would be dominated by the Sun and sources quite close to the Solar system due to the strong absorption \citep{Jester2009NewAR,cong_ultralong-wavelength_2021}. To avoid complicating the problem, here we limit our investigation in the frequency of 1 - 30 MHz. 
We re-bin the TOD to generate the mock data sample. Then we parameterize the spectrum, using a turnover frequency to describe the absorption. Finally, we divide the sky into multiple regions and reconstruct the parameterized spectrum for each region from the mock data sample.

\subsection{The mock observational data sample}

In our simulation setup, the Moon is a sphere with a radius of 1737.47 km. The satellite moves around the Moon in a circular orbit at a constant height of 300 km from the Moon surface. The period of one orbit  
is about 2.3 hours. According to the DSL design, the orbital plane of the satellite has an inclination angle of $30^\circ$ with respect to the lunar equator plane, and the orbital plane precesses relative to the lunar equator plane with a period of 1.3 years \citep{chen_discovering_2021}. In this study, we use only one of the antennas, and assume that the antenna always points to the anti-lunar center and that the maximum of the beam response is located in the plane perpendicular to the antenna. When it moves along the orbit, the antenna pointing changes with time, and the Moon blocks the sky area in the opposite direction of the pointing.

Suppose that at time $t$ the antenna is pointing in the direction $\hat{\boldsymbol{k}}_t$ on the sky, then the antenna temperature at frequency $\nu$ is \citep{shi_lunar_2022},
\begin{align}
    & T_{\rm TOD}(\nu,t) \nonumber\\
    &= \frac{\int S(\hat{\boldsymbol{k}},\hat{\boldsymbol{k}}_t)B(\hat{\boldsymbol{k}},\hat{\boldsymbol{k}}_t)T_{\rm sky}(\nu,\hat{\boldsymbol{k}})d\Omega(\hat{\boldsymbol{k}})}
    {\int B(\hat{\boldsymbol{k}},\hat{\boldsymbol{k}}_t)d\Omega(\hat{\boldsymbol{k}})}
     + T_{\rm N}(\nu, t).
\label{eq:T_a}
\end{align}
Here $T_{\rm sky}$ is the input sky brightness temperature using the {\tt ULSA} model, $S(\hat{\boldsymbol{k}},\hat{\boldsymbol{k}}_t)$ is the shade function, which describes the lunar occultation of sky, $B(\hat{\boldsymbol{k}},\hat{\boldsymbol{k}}_t)$ is the antenna primary beam response, and $T_{\rm N}(\nu, t)$ is the thermal noise. 
If radiation from direction $\hat{\boldsymbol{k}}$ is blocked by the Moon, $S(\hat{\boldsymbol{k}},\hat{\boldsymbol{k}}_t) = 0$, otherwise $S(\hat{\boldsymbol{k}},\hat{\boldsymbol{k}}_t) = 1$. 
The total length of the antenna onboard each of the DSL low-frequency daughter satellites is 5 m, which is equal to half wavelength at 30 MHz and smaller than half wavelength at all frequencies below 30 MHz. Therefore, the beam response can be well approximated as a short dipole \citep{balanis2016antenna}, i.e.
\begin{equation}
B(\hat{\boldsymbol{k}},\hat{\boldsymbol{k}}_t) = 1 - (\hat{\boldsymbol{k}}\cdot\hat{\boldsymbol{k}}_t)^2.
\end{equation}
While the beam chromaticity is a very important issue for experiments aiming at extracting the weak cosmic 21-cm signals (e.g. \citealt{Vedantham,Bernardi,Mozdzen,Thyagarajan,Spinelli,Mahesh,Anstey2021,Kim}), here we focus on the foreground sky itself, especially the absorption features at low-frequencies, the chromatic beam effect is negligible as compared to the reconstruction errors in the spectrum (see results in Section~\ref{sec:results}).
For such a beam response, the normalization factor in the denominator of Eq. (\ref{eq:T_a}) equals  $8\pi/3$.

\begin{figure}[ht]
  \centering
  \includegraphics[width=0.4\textwidth]{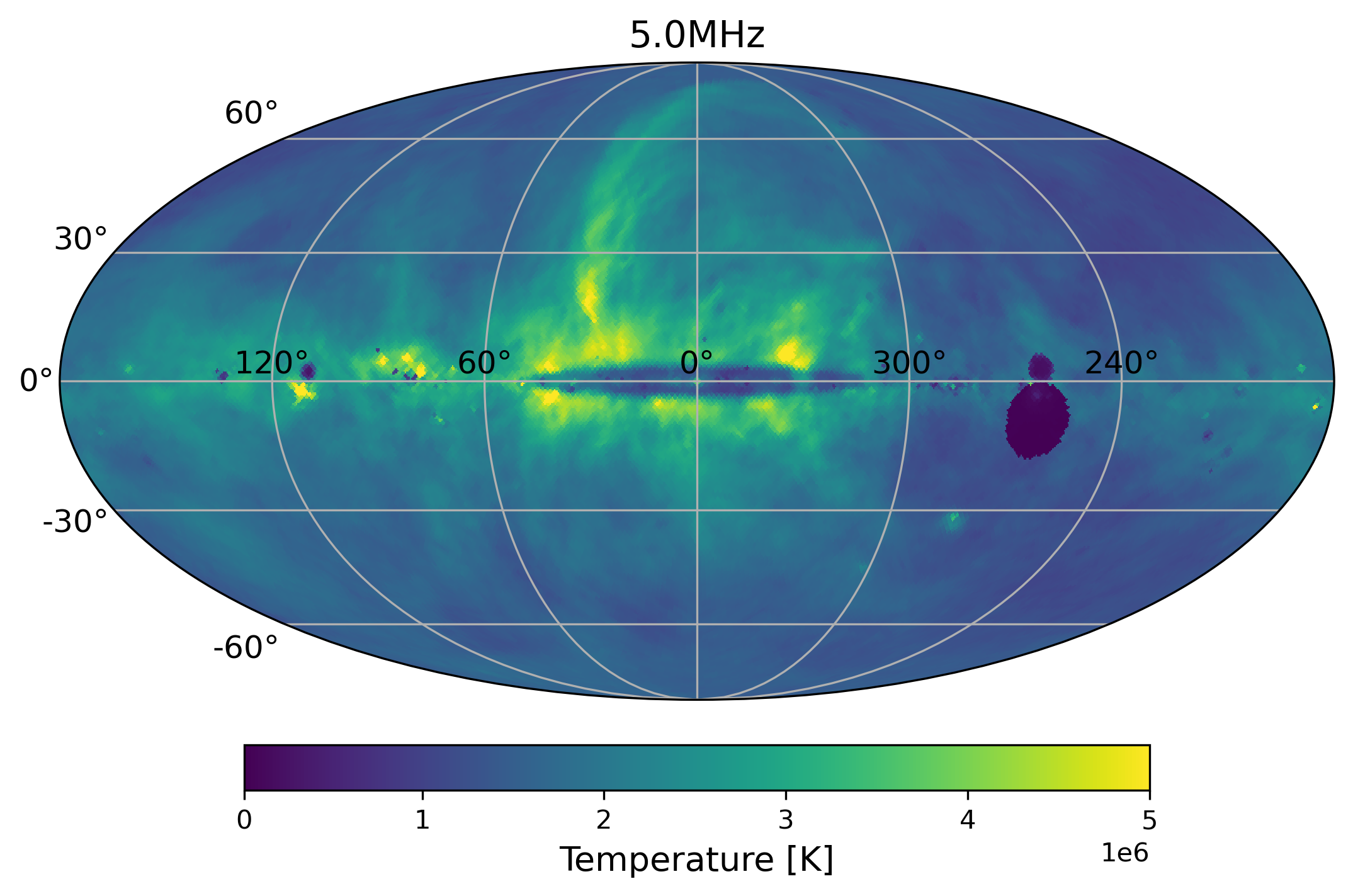}
  \hspace{0.5cm}  
  \caption{The sky map with frequency of 5 MHz in the Galactic coordinate system. 
  }
  \label{fig:5MHz_map}
\end{figure}

The input sky map adopts $N_{\rm side} = 64$ in the HEALPix scheme \citep{gorski_healpix_2005}, corresponding to a resolution of about 55 arcmin, which is high enough for the following low-resolution reconstruction. In Fig. \ref{fig:5MHz_map} we plot the input sky map at 5 MHz. The map shows clear free-free absorption features: the central Galactic plane is darker than surrounding areas, there are some dark spots near the Galactic plane which are dense H II regions.   

The thermal noise $T_{\rm N}(\nu, t)$ is assumed to follow a Gaussian distribution with zero mean and standard deviation being
\begin{equation}
   \sigma_{\rm N} (t) = \frac{T_{\rm sys}(t)}{\sqrt{\Delta\nu t_{\rm int}}},
\label{eq:noise}
\end{equation}
where $t_{\rm int}$ is the integration time of each data point, and $\Delta \nu$ is the width of each frequency point. The system temperature $T_{\rm sys}(t)$ has contributions from both sky radiation and receiver noise, 
\begin{equation}
    T_{\rm sys}(t) = T_{\rm rcv} + \bar{T}_{\rm sky}(t).
\end{equation}
$\bar{T}_{\rm sky}(t)$ is the mean brightness temperature of the sky as averaged over the antenna beam that points in the direction of
$\hat{\boldsymbol{k}}_t$. $T_{\rm rcv}$ is the effective temperature of the receiver noise. It can be approximated as \citep{shi_imaging_2022}
\begin{equation}
    T_{\rm rcv} = 6.61 \times 10^3 \left(\frac{\lambda}{\rm 10~m}\right)^2 
    \left(\frac{\rm 2.5~m}{l_{
    \rm eff}}\right)^2~[\rm K],
\end{equation}
where $l_{\rm eff}$ is the effective length of the antenna wire, which is assumed to be 2.5 m here.

One precession period of DSL lasts 1.3 yr. If we sample an antenna temperature every 1 second, there will be 40,996,800 data points for each frequency. One way to reduce the number of data points is to increase the integration time. However, from Eq. (\ref{eq:T_a}), the antenna temperature varies rapidly as the Moon blockage changes when the antenna pointing direction  $\hat{\boldsymbol{k}}_t$ varies over time. Spatial information would be lost if the integration is too long. Even if we sample an antenna temperature point every 300 seconds to ensure a low-resolution reconstruction, then for each frequency there will still be 136,656 data points. It is a huge load for the later reconstruction procedure.

To reduce computational cost, 
we re-pixelize the TOD with the HEALPix scheme \citep{gorski_healpix_2005} and adopt $N_{\rm side}=8$ (corresponding to a resolution $439.7$ arcmin), then average the TOD points whose antenna pointings are in the same pixel.  
Then the $i$-th data point in the mock data sample is 
\begin{equation}
    T_{{\rm data},i}(\nu) = \frac{1}{N_i}\sum_{h=1}^{N_i}T_{{\rm TOD},i,h}(\nu), \quad 
\label{eq:T_datai}
\end{equation}
where $N_i$ is the number of TOD points in the $i$-th pixel and $T_{{\rm TOD},i,h}$ is its $h$-th TOD point at frequency $\nu$.

\begin{figure}[ht]
  \centering
\includegraphics[width=0.4\textwidth]{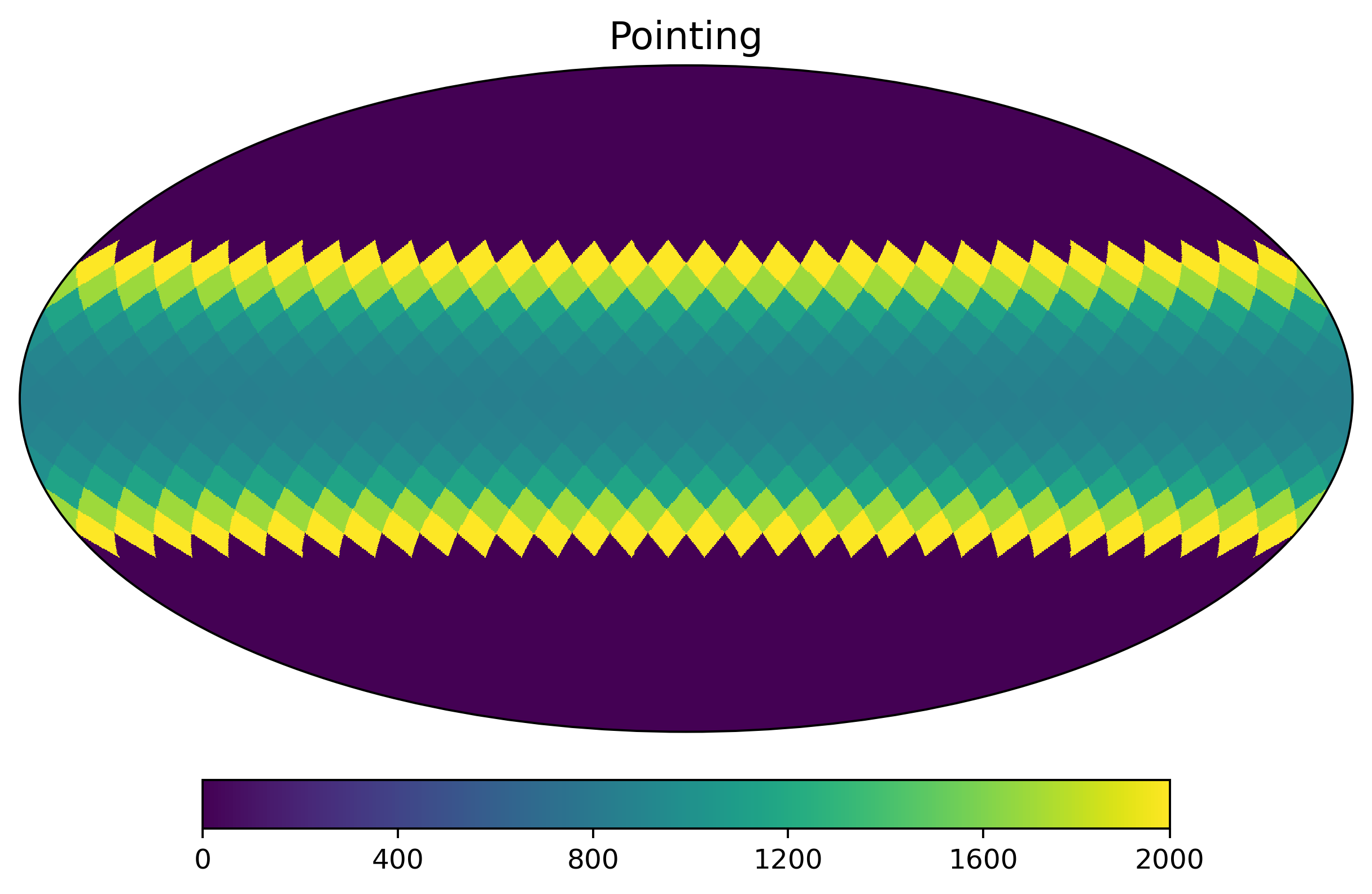}
  \caption{
  In one precession period, the antenna wire repeatedly points into same pixels on sky. This figure shows the number of pointing times for each  sky pixel in {\it lunar equatorial coordinate system}. The sky is pixelized by HEALPix scheme with $N_{\rm side}=8$.  
  }
 \label{fig:points}  
\end{figure}

In Fig. \ref{fig:points} we plot the number of data points in each pixel.  It illustrates that the antenna fails to point in any area with lunar declination $> 30^\circ$ or $<-30^\circ$, because the orbit of the DSL has an inclination angle of $30^\circ$ and the antenna is assumed to always point away from the Moon center. However, we note that each pixel just represents the central pointing direction of the antenna wire. The dipole antenna has a very wide beam coverage, and radiation from regions $>30^\circ$ far from the lunar equator is still received by the antenna.

\begin{figure}[ht]
  \centering
\includegraphics[width=0.4\textwidth]{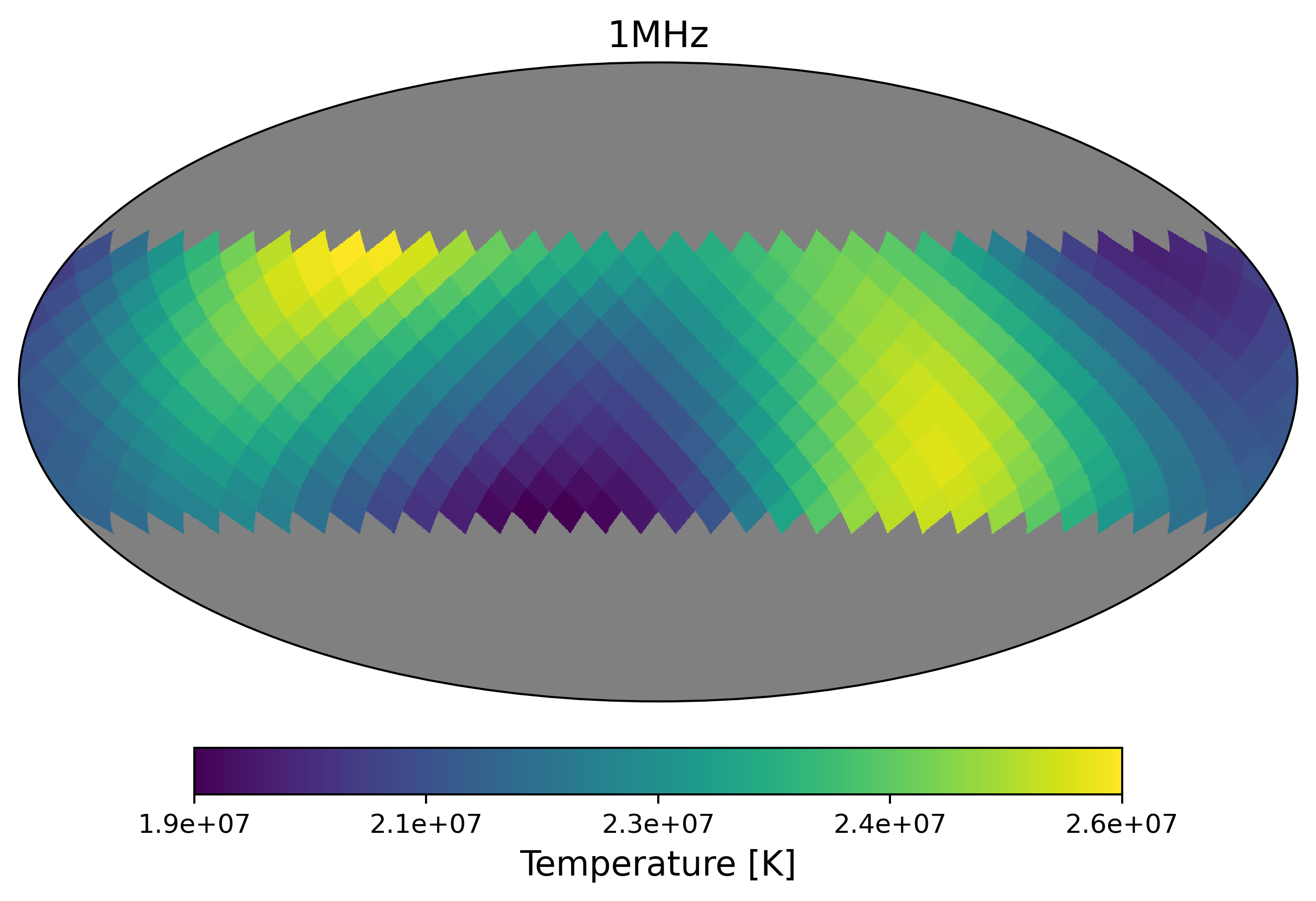}
  \caption{ 
  Our mock data sample for 416 pixels on the sky in the lunar equatorial coordinate system. Note that this is not the sky temperature distribution, but the mean antenna temperature when antenna wire is pointing into the corresponding pixel. 
  }
 \label{fig:mock_sample}  
\end{figure}

After the above procedure, the TOD is reduced to mock observation data sample 
with 416 data points for each frequency. We plot the mock data sample on the sky according to the antenna pointing in Fig. \ref{fig:mock_sample}. Please note that this figure shows the averaged antenna temperature when the antenna points to the pixel. It is the full-sky temperature weighted by the primary beam response and Moon blockage, not the sky temperature from that pixel.

\subsection{Reconstruction of parameterized spectrum from mock data sample}

\begin{figure}[ht]
  \centering  \includegraphics[width=0.4\textwidth]{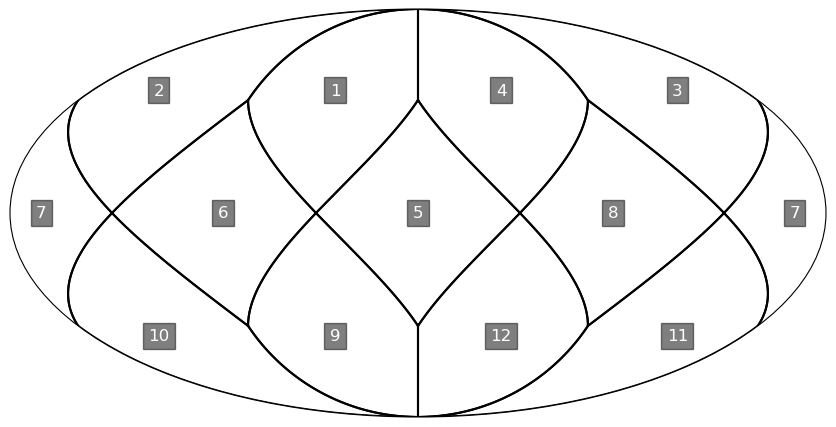}
 \caption{ 
 The position of 12 regions in the Galactic coordinate system. The sequence of each region is marked in the figure. 
 }
  \label{fig:regions}
\end{figure}

To reconstruct the spectra of different sky regions, we divide the sky into $N_{\rm region}=12$ regions (corresponding to $N_{\rm side}=1$ in the HEALPix scheme); their positions in the sky in the Galactic coordinate system are shown in Fig. \ref{fig:regions}. 
In principle, we can arbitrarily divide the sky into any number of regions and each of them could have a different area. However, since a single dipole antenna has a very wide primary beam, when coupled with the large-scale Moon blockage, it is inadequate for resolving small-scale sky temperature anisotropies. If the number of regions is too large, the reconstructed spectra for neighboring regions would be highly degenerate. Therefore, in this work, we only focus on large-scale anisotropies of the ultra-long wavelength spectrum, and adopt a modest choice of $N_{\rm region}=12$.

We model the $i$-th mock data point  
as the weighted sum of contribution from all the 12 sky regions,
\begin{align}
&T_{{\rm model},i}(\nu) 
& \\
&= \frac{1}{(8\pi/3)} \sum_{j=1}^{N_{\rm region}} \int_{\Delta \Omega_j}S(\hat{\boldsymbol{k}} , \hat{\boldsymbol{k}}_i)B(\hat{\boldsymbol{k}} ,\hat{\boldsymbol{k}}_i)T^{\rm m}_{{\rm sky},j}(\nu)\, d\Omega(\hat{\boldsymbol{k}}) \nonumber \\
&= \frac{1}{(8\pi/3)}    
\sum_{j=1}^{N_{\rm region}}  \bar{\mathcal{B}}_j(\hat{\boldsymbol{k}}_i)T^{\rm m}_{{\rm sky},j}(\nu)\Delta \Omega_j,
\label{eq:Ti}
\end{align}

where $\hat{\boldsymbol{k}
}_i$ is the direction of the $i$-th point in the modeled data sample, 
$T^{\rm m}_{{\rm sky},j}$ is the average sky temperature of the $j$-th region, and   $\triangle\Omega_j$= $4\pi/N_{\rm region}$ is the area of each region.
\begin{equation}
\mathcal{\bar{B}}_j(\hat{\boldsymbol{k}}_i)=\frac{1}{\Delta \Omega_j}\int_{\Delta\Omega_j} S(\hat{\boldsymbol{k}} , \hat{\boldsymbol{k}}_i)B(\hat{\boldsymbol{k}} ,\hat{\boldsymbol{k}}_i)d\Omega(\hat{\boldsymbol{k}})
\end{equation}
is the mean beam response weighted by the shade function for the $j$-th region.

The parameterization
of $T^{\rm m}_{{\rm sky},j}$ is the target we are going to reconstruct.
We use the smoothly broken power law to parameterize the mean spectra in each region, implemented via Astropy's \texttt{SmoothlyBrokenPowerLaw1D} \citep{astropy2013,astropy2018}. For the $j$-th region of the sky, the spectrum is modeled as
\begin{align}
& T^{\rm m}_{{\rm sky},j}(\nu) \\
&= A_j \left( \frac{\nu}{\nu_{t,j}} \right)^{-a_j} 
\left[ \frac{1}{2} \left( 1 + \left( \frac{\nu}{\nu_{t,j}} \right)^{1/\Delta_j} \right) \right]^{(a_j - b_j) \Delta_j},
\label{eq:fre_spectrum}
\end{align} 
where $A_j$, $\nu_{t,j}$, $a_j$, $b_j$ and $\Delta_j$ are the five parameters of each region.
The parameter $\nu_{t,j}$ corresponds to the critical frequency characterizing a turnover feature, possibly caused by free-free absorption. For $\nu\gg\nu_{t,j}$, the spectral index is approximately $-b_j$, 
and for $\nu\ll\nu_{t,j}$, the spectral index $\thickapprox -a_j$. $\Delta_j$ controls the smoothness of the transition between the two end slopes. 
However, due to the  complicated distribution of emissivity and electron distribution in the Milky Way, the average spectrum of a region cannot be perfectly fitted with the above formula. We have checked that, for the 12 regions, at $\nu \gtrsim 1$ MHz, their frequency spectrum can be fitted by Eq. (\ref{eq:fre_spectrum}) with errors of percentage level. 
 
We then perform reconstruction of the frequency spectrum for the 12 regions from the mock data sample, using the Markov Chain Monte Carlo (MCMC) method with {\tt EMCEE} \citep{Foreman-Mackey_2013}. The spectrum of each region is described by  five parameters, so there are 60 parameters in total.
The likelihood function
\begin{equation}
L(\boldsymbol{\theta}|\boldsymbol{x}) \propto \exp\left(-\frac{1}{2} (\mathbf{x} - \boldsymbol{\mu}(\boldsymbol{\theta}))^\intercal \boldsymbol{\Sigma}^{-1} (\mathbf{x} - \boldsymbol{\mu}(\boldsymbol{\theta}))\right),
\label{eq:MCMC}
\end{equation}
where $\mathbf{x}$ is the mock data set,  $\boldsymbol{\mu}(\boldsymbol{\theta})$ is the modeled antenna temperature given by Eq. (\ref{eq:Ti}), corresponding to the parameter set $\boldsymbol{\theta}$, and $\boldsymbol{\Sigma}$ is the covariance matrix containing errors that will be introduced in the next subsection.  Throughout this paper, we adopt 59 frequency points equally centered from 1 to 30 MHz, each has width $\Delta \nu = 0.5 $ MHz. Since the TOD sample at each frequency is grouped into 416 pixels, we have $416\times 59=24544$ mock data points in total.

\subsection{The errors}
\label{sec:errors}
In addition to thermal noise, our pipeline contains three primary sources of errors, all of which should be included in the covariance matrix.
The first error comes from the averaging process, in which TOD points are grouped in each of the downgraded pixel to generate the mock data sample, while the modeled data sample is generated assuming that the antenna points to the pixel center. This averaging procedure results in a deviation of the modeled data from the mock data and causes the {\it averaging error} which is $\lesssim 0.1\%$.
The second error arises from the limited number of sky regions in the spectrum reconstruction process, and for each region an average spectrum is used to represent the spectra of an entire region. This is called the {\it discretization error}, and results in deviations of the modeled data that range from approximately $\sim 1\%$ to $\sim 7\%$ of the mock data. The third is the {\it model error} caused by the inaccuracy of applying Eq. (\ref{eq:fre_spectrum}) to the average spectra. For frequencies $\gtrsim 10$ MHz, this error is $\sim 1\%$, and for frequencies between $\sim 10$ and $ \sim 1$ MHz, this error increases linearly from $\sim 1\%$ to $\sim 6\%$.

First, we calculate the averaging error in the mock data itself. When generating the mock data sample, each data is obtained by averaging TOD points for which the antenna pointing directions fall in the same pixel of 439.7 arcmin resolution ($N_{\rm side}=8$). The values of TOD points in the same pixel are not exactly identical. Due to the limited number of TOD points and non-uniform distribution of pointings, such data-averaging scheme causes a deviation between the average value of TOD (the mock data) and the antenna temperature calculated under the assumption that the antenna points to the pixel center (the model). Of course this error vanishes if we have an infinite number of TOD points uniformly distributed in each pixel. We write the covariance matrix of averaging error as
\begin{widetext}
\begin{equation}
 C^a_{ii'}(\nu_f, \nu_{f'})=
\begin{cases}
    \frac{N_i - 1}{N_i} \sum_{h=1}^{N_i} (T_{{\rm data},ih}(\nu_f) - T_{{\rm data},i}(\nu_f)) (T_{{\rm data},ih}(\nu_{f'}) - T_{{\rm data},i}(\nu_{f'})), ~&i=i'\\
    0,~&i\neq i',
    \end{cases}
    \label{eq:C_ii2}
\end{equation}
where $T_{{\rm data},i}$ is given by Eq. 
(\ref{eq:T_datai}), and
\begin{equation}
    T_{{\rm data},ih}(\nu_f) = \frac{\sum_{r=1,r\neq h}^{N_i}T_{{\rm TOD},i,r}(\nu_f)}{N_i-1}
\end{equation}
represents the average value of TOD data points in the $i$-th pixel after excluding the $h$-th one, at frequency $\nu_f$. The $i$-$i'$ correlation term of $C^a_{ii'}(\nu_f, \nu_{f'})$ actually represents the Jackknife error \citep{jackknife_1,jackknife_2}.
As the TOD include the thermal noise, this averaging error naturally includes the error due to thermal noise.
\end{widetext}

\begin{figure}[ht]
  \centering
\includegraphics[width=0.4\textwidth]{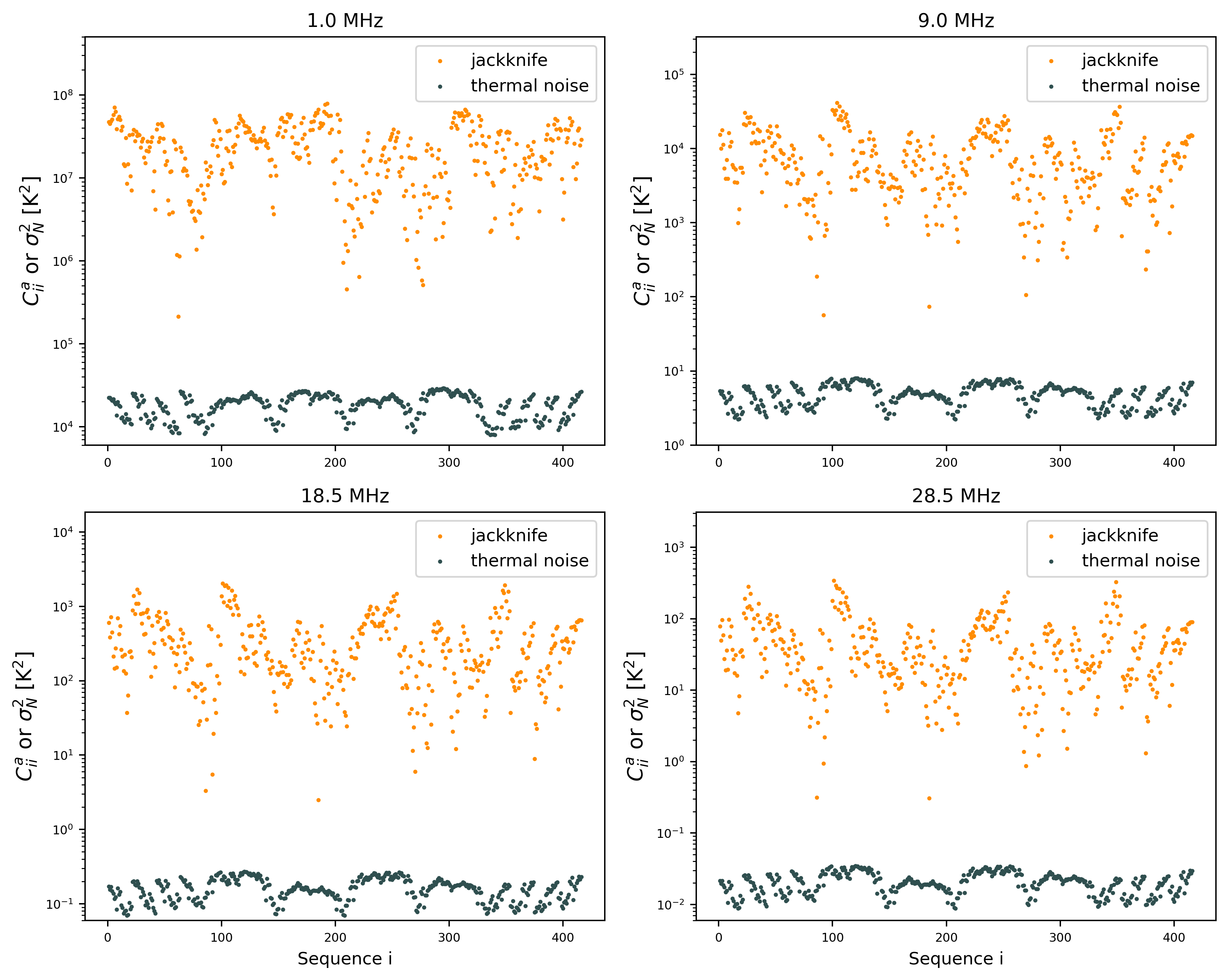}
  \caption{ 
  The averaging errors described by Eq. (\ref{eq:C_ii2}) compared with thermal noise.
  }
 \label{fig:jack_thermal}  
\end{figure}

In Fig. \ref{fig:jack_thermal} we plot $C^a_{ii}$, compared with thermal noise, at four different frequencies. The wide spread between the Jackknife error and the thermal noise indicates that the averaging error is dominated by the sample variance in the averaging process.

\begin{figure}[h]
  \centering
\includegraphics[width=0.4\textwidth]{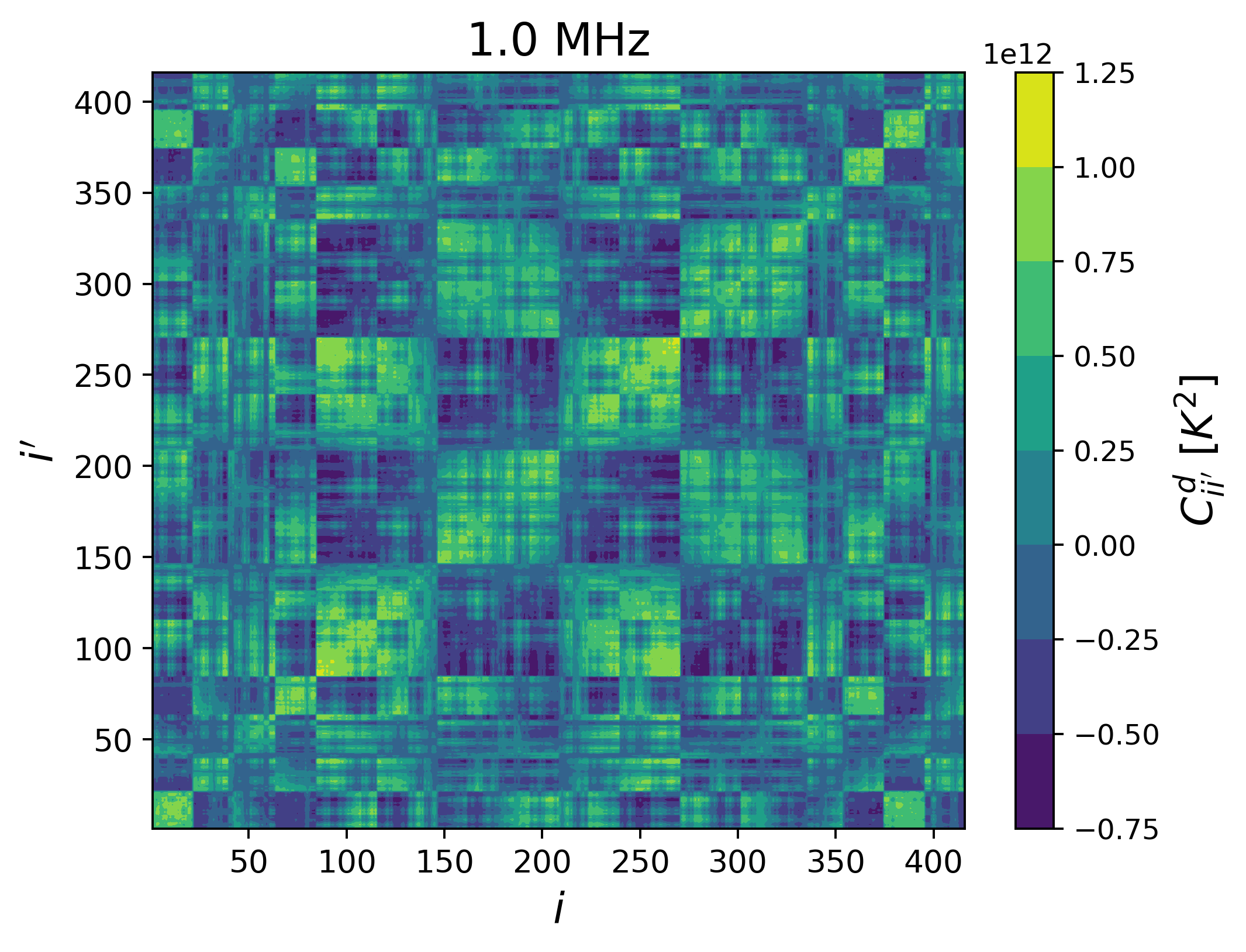}
  \caption{ 
    The covariance ${C^d_{ii'}}$ expressed by Eq. (\ref{eq:C_ii3}),  for $\nu_f=\nu_{f'}=1$ MHz.
  }
 \label{fig:C_i1i2}  
\end{figure}

The averaging error will of course depend on the size of the pixel, and a higher resolution could, in principle, improve precision. However, the size of the covariance matrix increases quadratically with the sample size, and a higher resolution would lead to a larger sample. 
With the adopted resolution of $N_{\rm side}=8$, the relative error introduced by averaging data points is subdominant compared to  errors introduced later on.
Therefore, $N_{\rm side}=8$ provides a good balance between accuracy and computational feasibility, and choosing a higher resolution would not influence the results.

Another two errors of modeling the data with Eq. (\ref{eq:Ti}) arise from the limited number ($N_{\rm region} = 12$) of sky regions in the reconstruction process (discretization error), and the inaccuracy of the spectrum model Eq. (\ref{eq:fre_spectrum}) in describing the average brightness temperature for a region (model error). They could both cause the deviation $\delta T$ between mock data sample and the model. 
Excluding the averaging error, the difference between $T_{{\rm data},i}$ and $T_{{\rm model},i}$ is
\begin{align}
&\delta T_i (\nu_f) \nonumber \\
&= \frac{1}{(8\pi/3)} \sum_{j=1}^{N_{\rm region}} \left[\int_{\Delta \Omega_j} S(\hat{\boldsymbol{k}} , \hat{\boldsymbol{k}}_i)B(\hat{\boldsymbol{k}} ,\hat{\boldsymbol{k}}_i)T_{\rm sky}(\nu_f, \hat{\boldsymbol{k}})d\Omega(\hat{\boldsymbol{k}})\right. \nonumber \\
&\quad \left. - \mathcal{\bar{B}}_j(\hat{\boldsymbol{k}}_i) T^{\rm m}_{{\rm sky},j}(\nu_f)  \Delta \Omega_j\right] .
\end{align}

\begin{widetext}
To separate the discretization error from the model error, we introduce
the mean sky temperature for each region of the sky $\bar{T}_{{\rm sky},j}$, which can be expressed as
\begin{equation}
\bar{T}_{{\rm sky},j}(\nu_f) = \frac{1}{\Delta \Omega_j}\int_{\Delta \Omega_j}T_{\rm sky}(\nu_f,\hat{\boldsymbol{k}})d\Omega(\hat{\boldsymbol{k}}),
\label{eq:aveT_j}
\end{equation}
and transform $\delta T_i$ into the following form (for concision we omit the variable $\nu_f$ in this derivation)
\begin{align}
\delta T_i 
&= \frac{1}{(8\pi/3) } \sum_{j=1}^{\rm N_{\rm region}}\left[\int_{\Delta \Omega_j}
S(\hat{\boldsymbol{k}} , \hat{\boldsymbol{k}}_i)B(\hat{\boldsymbol{k}} ,\hat{\boldsymbol{k}}_i)T_{\rm sky}(\hat{\boldsymbol{k}})d\Omega(\hat{\boldsymbol{k}}) 
-\mathcal{\bar{B}}_j(\hat{\boldsymbol{k}}_i) T^{\rm m}_{{\rm sky},j}  \Delta \Omega_j+\mathcal{\bar{B}}_j(\hat{\boldsymbol{k}}_i) \bar{T}_{{\rm sky},j}  \Delta \Omega_j-\mathcal{\bar{B}}_j(\hat{\boldsymbol{k}}_i) \bar{T}_{{\rm sky},j}  \Delta \Omega_j
\right] \nonumber \\
&= \frac{1}{(8\pi/3) } \sum_{j=1}^{\rm N_{\rm region}}\left[\int_{\Delta \Omega_j}
[S(\hat{\boldsymbol{k}} , \hat{\boldsymbol{k}}_i)B(\hat{\boldsymbol{k}} ,\hat{\boldsymbol{k}}_i)-\mathcal{\bar{B}}_j(\hat{\boldsymbol{k}}_i)]T_{\rm sky}(\hat{\boldsymbol{k}})d\Omega(\hat{\boldsymbol{k}})\right]+
\frac{1}{(8\pi/3) }\sum_{j=1}^{\rm N_{\rm region}}
(\bar{T}_{{\rm sky},j}-T^{\rm m}_{{\rm sky},j})\mathcal{\bar{B}}_j(\hat{\boldsymbol{k}}_i)\Delta \Omega_j \nonumber \\
&= \frac{1}{(8\pi/3) }  \left[\int_{ 4\pi}
[S(\hat{\boldsymbol{k}} , \hat{\boldsymbol{k}}_i)B(\hat{\boldsymbol{k}} ,\hat{\boldsymbol{k}}_i)-\mathcal{\bar{B}}(\hat{\boldsymbol{k}},\hat{\boldsymbol{k}}_i)]T_{\rm sky}(\hat{\boldsymbol{k}})d\Omega(\hat{\boldsymbol{k}})\right]+
\frac{1}{(8\pi/3) }\sum_{j=1}^{\rm N_{\rm region}}
(\bar{T}_{{\rm sky},j}-T^{\rm m}_{{\rm sky},j})\mathcal{\bar{B}}_j(\hat{\boldsymbol{k}}_i)\Delta \Omega_j.
\label{eq:deltaT_i}
\end{align} 
\end{widetext}
Here the second line follows when combining the first and fourth terms in the first line and substituting $\bar{T}_{{\rm sky},j}$ with the integral of Eq.(\ref{eq:aveT_j}). 
In the last line of Eq. (\ref{eq:deltaT_i}), the integral is over the whole sky, and $\mathcal{\bar{B}}(\hat{\boldsymbol{k}}, \hat{\boldsymbol{k}}_i)$ is defined as the full-sky mosaic of $\mathcal{\bar{B}}_j(\hat{\boldsymbol{k}}_i)$, i.e. $\mathcal{\bar{B}}(\hat{\boldsymbol{k}}, \hat{\boldsymbol{k}}_i)=\mathcal{\bar{B}}_j(\hat{\boldsymbol{k}}_i)$ if $\hat{\boldsymbol{k}}$ falls in region $j$.
The first term of this line describes the discretization error, and the second term represents the model error.

\begin{widetext}
As long as we exactly know the beam, we can calculate the discretization error, and incorporate it into the covariance matrix.
We further define a differential beam
\begin{equation}
\delta \mathcal{{B}}(\hat{\boldsymbol{k}},\hat{\boldsymbol{k}}_i) = S(\hat{\boldsymbol{k}} , \hat{\boldsymbol{k}}_i)B(\hat{\boldsymbol{k}} ,\hat{\boldsymbol{k}}_i)-\mathcal{\bar{B}}(\hat{\boldsymbol{k}},\hat{\boldsymbol{k}}_i),
\end{equation}
then the first term of Eq. (\ref{eq:deltaT_i}) 
can be written in spherical harmonic space as
\begin{align}
&\frac{1}{(8\pi/3) }  \int_{ 4\pi}
\delta \mathcal{{B}}(\hat{\boldsymbol{k}},\hat{\boldsymbol{k}}_i)T_{\rm sky}(\hat{\boldsymbol{k}})d\Omega(\hat{\boldsymbol{k}})\nonumber\\
=&
\frac{1}{(8\pi/3) }\int_{ 4\pi}
\left(\sum_{l'=0}^{l_{\rm max}}\sum_{m'=-l'}^{l'} 
\delta \mathcal{\tilde{B}}^*_{l'm'}(\hat{\boldsymbol{k}}_i)
Y^*_{l'm'}(\hat{\boldsymbol{k}}) \right)
\left(\sum_{l=0}^{l_{\rm max}}\sum_{m=-l}^l 
\tilde{T}_{{\rm sky},lm}  Y_{lm}(\hat{\boldsymbol{k}})\right)
d\Omega(\hat{\boldsymbol{k}})\nonumber\\
=& \frac{1}{(8\pi/3) }\int_{ 4\pi}
\sum_{l=0}^{l_{\rm max}}\sum_{m=-l}^l \sum_{l'=0}^{l_{\rm max}}\sum_{m'=-l'}^{l'} \tilde{T}_{{\rm sky},lm} \delta \mathcal{\tilde{B}}^*_{l'm'}(\hat{\boldsymbol{k}}_i) Y_{lm}(\hat{\boldsymbol{k}}) Y^*_{l'm'}(\hat{\boldsymbol{k}})d\Omega(\hat{\boldsymbol{k}})\nonumber\\
=& \frac{1}{(8\pi/3) }
\sum_{l=0}^{l_{\rm max}}\sum_{m=-l}^l \tilde{T}_{{\rm sky},lm} \delta \mathcal{\tilde{B}}^*_{lm}(\hat{\boldsymbol{k}}_i), 
\label{eq:deltaTi_1_sh_transform}
\end{align}
\end{widetext}

where $\delta \mathcal{\tilde{B}}_{lm}(\hat{\boldsymbol{k}}_i)$ and $\tilde{T}_{{\rm sky},lm}$ are spherical harmonic transforms of $\delta \mathcal{B}(\hat{\boldsymbol{k}},\hat{\boldsymbol{k}}_i)$ and $T_{\rm sky}(\hat{\boldsymbol{k}})$ respectively,
and the second line of Eq.~(\ref{eq:deltaTi_1_sh_transform}) follows because $\delta \mathcal{{B}}(\hat{\boldsymbol{k}},\hat{\boldsymbol{k}}_i)$ is real. In the following calculation, we adopt $l_{\rm max} = 128$ which is found to be sufficient to represent $\delta \mathcal{B}(\hat{\boldsymbol{k}},\hat{\boldsymbol{k}}_i)$. 
\begin{widetext}
The contribution of the first term in Eq. (\ref{eq:deltaT_i}) to the covariance matrix is
\begin{align}
C^d_{ii'} (\nu_f, \nu_{f'}) 
&= \frac{1}{(8\pi/3)^2}\left\langle \left(\sum_{l,m} 
\delta \mathcal{\tilde{B}}^*_{lm}(i)\tilde{T}_{{\rm sky},lm}(\nu_f) \right)*\left(\sum_{l',m'} 
\delta \mathcal{\tilde{B}}^*_{l'm'}(i')\tilde{T}_{{\rm sky},l'm'} (\nu_{f'}) \right)^* \right\rangle \nonumber \\
&=\frac{1}{(8\pi/3)^2} \sum_{l,m} \sum_{l',m'} 
\delta \mathcal{\tilde{B}}^*_{lm}(i) \delta \mathcal{\tilde{B}}_{l'm'}(i')
\langle \tilde{T}_{{\rm sky},lm}(\nu_f)
\tilde{T}^*_{{\rm sky},l'm'}(\nu_{f'}) \rangle.
\end{align}
\end{widetext}

Here $\langle \rangle$ is to take the expected value w.r.t the unknown sky temperature distribution.
We assume that different $(l, m)$ modes are independent, then the covariance matrix due to discretization error can be written as
{\small \begin{align}
&C^d_{ii'} (\nu_f, \nu_{f'})  \\
&= \frac{1}{(8\pi/3)^2}
\sum_{lm}\sum_{l'm'}
\delta \mathcal{\tilde{B}}^*_{lm}(i)
\delta \mathcal{\tilde{B}}_{l'm'}(i') \nonumber \\
&\quad * \left\langle \tilde{T}_{{\rm sky},lm}(\nu_f) \tilde{T}^*_{{\rm sky},lm}(\nu_{f'}) \right\rangle \delta_{ll'}\delta_{mm'} \\
&= \frac{1}{(8\pi/3)^2} \sum_{lm} \delta \mathcal{\tilde{B}}^*_{lm}(i)
\delta \mathcal{\tilde{B}}_{lm}(i') 
\left\langle \tilde{T}_{{\rm sky},lm}(\nu_f) \tilde{T}^*_{{\rm sky},lm}(\nu_{f'}) \right\rangle,
\end{align}}
where $\delta_{ll'}$ and $\delta_{mm'}$ are Kronecker deltas. For isotropic distribution of sky temperature fluctuations, $\langle \tilde{T}_{{\rm sky},lm}(\nu_f) \tilde{T}^*_{{\rm sky},lm}(\nu_{f'}) \rangle$
is independent of $m$, and actually equals to angular power spectrum, then the above formula can be further approximated as
\begin{align}
C^d_{ii'}(\nu_f, \nu_{f'})
= \frac{1}{(8\pi/3)^2} \sum_{lm} \delta \mathcal{\tilde{B}}^*_{lm}(i)
\delta \mathcal{\tilde{B}}_{lm}(i')C_l(\nu_f, \nu_{f'})
\label{eq:C_ii3}
\end{align} 
where $C_l(\nu_f, \nu_{f'}) = \langle \tilde{T}_{{\rm sky},lm}(\nu_f) \tilde{T}^*_{{\rm sky},lm}(\nu_{f'}) \rangle$ 
is the cross angular power spectrum between the sky temperature at frequency $\nu_f$ and $\nu_{f'}$. Note that the covariance matrix is not necessarily diagonal.
In Fig. \ref{fig:C_i1i2}, we plot the ${C^d_{ii'}}(\nu_f,\nu_f')$ at $\nu_f=\nu_{f'}= 1$ MHz in units of K$^2$. The mean brightness temperature of the sky increases toward lower frequencies, and is about  $2.6\times 10^7$ K at 1 MHz. 
It is found that $\sqrt{C^d_{ii}}$ accounts for approximately 0.65\% to 7\%  of the average sky temperature, which is the dominant source of error compared to the averaging error (Eq.(\ref{eq:C_ii2})), and is comparable to the model error estimated in Eq.(\ref{eq:C_ii4}). Moreover, the large amplitude of the nondiagonal elements of the matrix $C^d_{ii'}$ indicates that the different data points can be highly correlated. 

\begin{figure}[ht]
  \vspace{0.5cm} 
  \centering
\includegraphics[width=0.4\textwidth]{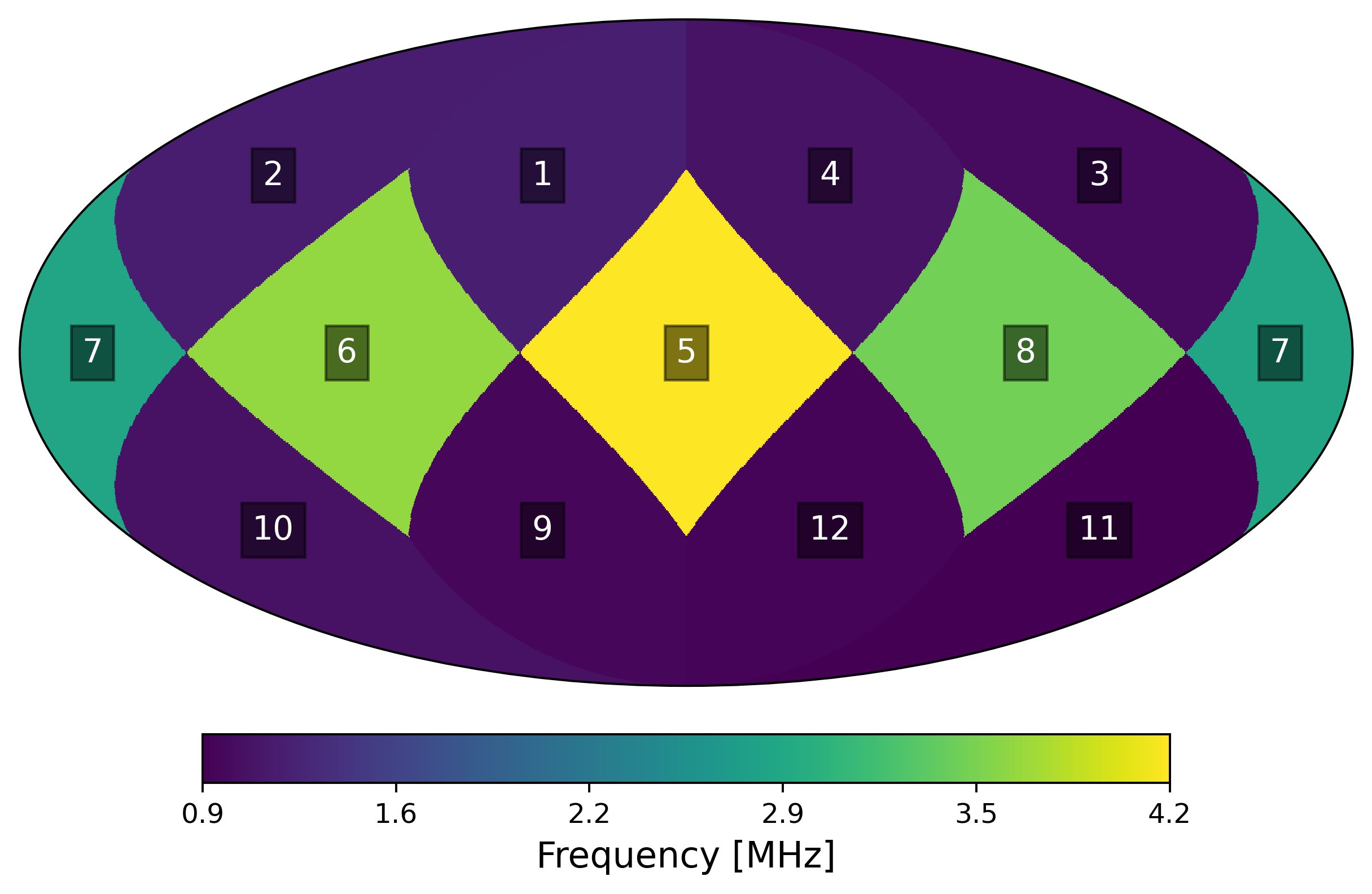}
  \caption{ 
  The turnover frequency $\nu_{t,j}$ for the 12 regions on sky in the Galactic coordinate system.}
 \label{fig:nu_j}  
\end{figure}

\begin{figure*}[htbp]
\centering
  \includegraphics[width=0.75\textwidth]{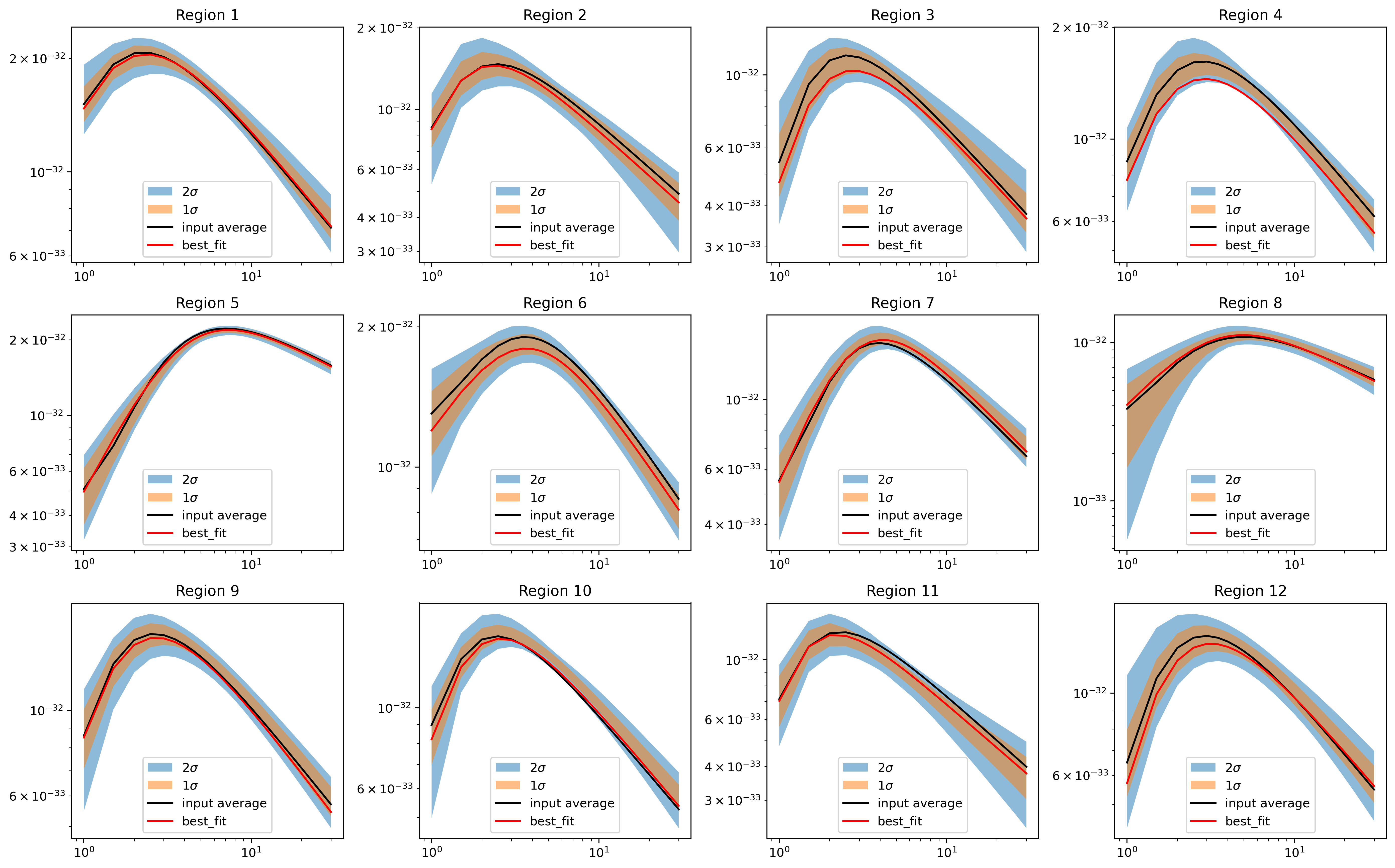}
  \caption{
  The fitted spectra for the 12 regions. To highlight the turnover feature on the spectra, the brightness temperature is converted into the specific intensity via the Rayleigh-Jeans law. In each panel, the black line is the mean intensity of the region from {\tt ULSA} sky model, the red line is the spectrum evaluated with the median values of the fitted parameters. The orange and blue shaded regions indicate the 1$\sigma$ and 2$\sigma$ uncertainties, respectively. 
  }
  \label{fig:mcmc_results}
\end{figure*}

The second term of Eq. (\ref{eq:deltaT_i}) is the model error when using Eq. (\ref{eq:fre_spectrum}) to fit the mean spectrum of each region in the sky. 
By directly fitting the spectra of pixels in the input {\tt ULSA} map using the spectrum model of Eq. (\ref{eq:fre_spectrum}), we find that the model error is $\sim 1\%$ for $\nu \gtrsim 10$ MHz, and gradually increases with decreasing frequency at $\lesssim 10$ MHz. At $\sim 1$ MHz the error reaches $\sim 6\%$.
\begin{widetext}
We therefore model the corresponding covariance as follows:
\begin{equation}
C^m_{ii'}(\nu_f, \nu_{f'})=
T_{{\rm data},i}(\nu_f) T_{{\rm data},i'}(\nu_{f'}) \mathcal{B}^m_{ii'} \times
\begin{cases}
\left(0.06 - 0.05 \frac{\ln(\nu_f/{\rm MHz})}{\ln 10} \right)^2, ~& \nu_f=\nu_{f'},\ \nu_f\leq 10{\rm \ MHz}\\
10^{-4}, ~& \nu_f=\nu_{f'},\ \nu_f>10{\rm \ MHz}\\
0, ~& \nu_f\neq\nu_{f'}
\end{cases},
\label{eq:C_ii4}
\end{equation}
\end{widetext}
where
\begin{equation}
    \mathcal{B}^m_{ii'} = \frac{ \int B(\hat{\boldsymbol{k}} ,\hat{\boldsymbol{k}}_i) B(\hat{\boldsymbol{k}} ,\hat{\boldsymbol{k}}_{i'}) d\hat{\boldsymbol{k}} }{\sqrt{\int B(\hat{\boldsymbol{k}} ,\hat{\boldsymbol{k}}_i) d\hat{\boldsymbol{k}} \int B(\hat{\boldsymbol{k}} ,\hat{\boldsymbol{k}}_{i'}) d\hat{\boldsymbol{k}}}}.
\end{equation}
This factor takes into account the correlation between different data samples due to possible overlapping beams.
Note that in Eq. (\ref{eq:C_ii4}) we neglect the correlation between different frequencies. In practice, the model error in different frequencies could be highly correlated if the spectrum was simply a power law as in the case of synchrotron radiation. However, in the ultra-long wavelength band with unknown free-free absorption distribution, modeling the correlation is rather complicated and requires additional information that is hard to obtain in real data analysis. By neglecting the frequency correlation in the model error, we introduce minimal prior to the reconstruction procedure, and the results are expected to be relatively conservative.

Finally, the full covariance matrix is the sum of $C^a$ (Eq. (\ref{eq:C_ii2})), $C^d$ (Eq. (\ref{eq:C_ii3})), and $C^m$ (Eq. (\ref{eq:C_ii4})).

\begin{table*}[htbp]
\centering
\begin{tabular}{|c|c|c|c|c|c|c|c|}
\hline
 & prior & $j=1$ & $j=2$ & $j=3$ & $j=4$ & $j=5$ & $j=6$   \\ 
\hline
  $(l,b)$ & \--- &$(45.0^\circ,41.8^\circ)$ &$(135.0^\circ,41.8^\circ)$&$(225.0^\circ,41.8^\circ)$&$(315.0^\circ,41.8^\circ)$&$(0.0^\circ,0.0^\circ)$&$(90.0^\circ,0.0^\circ)$\\
  \hline
 $\log A_j~[{\log\rm K}]$ & [5,20] & $17.48^{+0.28}_{-0.29}$ & $17.02^{+0.28}_{-0.27}$ & $16.55^{+0.28}_{-0.28}$ & $17.00^{+0.24}_{-0.25}$ & $15.11^{+0.21}_{-0.21}$ & $15.31^{+0.34}_{-0.33}$ \\
\hline
 $a_j$ & [-5,5] & $0.15^{+0.46}_{-0.60}$ & $-0.78^{+1.01}_{-1.46}$ & $-2.13^{+1.27}_{-1.42}$ & $-0.99^{+0.78}_{-1.21}$ & $0.71^{+0.23}_{-0.29}$ & $1.45^{+0.19}_{-0.25}$ \\
\hline
 $b_j$ & [-5,5] & $2.55^{+0.06}_{-0.05}$ & $2.56^{+0.10}_{-0.08}$ & $2.56^{+0.09}_{-0.08}$ & $2.57^{+0.06}_{-0.05}$ &
 $2.36^{+0.04}_{-0.03}$ &
 $2.53^{+0.07}_{-0.05}$ \\
\hline
 $\nu_{t,j}~[\rm MHz]$ & (0,30] & $1.18^{+0.25}_{-0.22}$ & $1.17^{+0.31}_{-0.31}$ & $1.00^{+0.32}_{-0.26}$ & $1.06^{+0.28}_{-0.24}$ & $4.16^{+0.57}_{-0.54}$ & $3.62^{+0.63}_{-0.57}$ \\
\hline
 $\Delta_j$ & (0,3] & $0.57^{+0.12}_{-0.11}$ & $0.43^{+0.16}_{-0.15}$ & $0.51^{+0.14}_{-0.13}$ & $0.60^{+0.13}_{-0.12}$ & $0.44^{+0.08}_{-0.08}$ & $0.44^{+0.15}_{-0.14}$ \\
\hline
 & prior & $j=7$ & $j=8$ & $j=9$ & $j=10$ & $j=11$ & $j=12$   \\ 
\hline
  $(l,b)$ & \--- &$(180.0^\circ,0.0^\circ)$ &$(270.0^\circ,0.0^\circ)$&$(45.0^\circ,-41.8^\circ)$&$(135.0^\circ,-41.8^\circ)$&$(225.0^\circ,-41.8^\circ)$&$(315.0^\circ,-41.8^\circ)$\\
  \hline
 $\log A_j~[\log\rm K]$ & [5,20] & $15.58^{+0.25}_{-0.31}$ & $14.87^{+0.34}_{-0.39}$ & $17.16^{+0.22}_{-0.21}$ & $17.07^{+0.22}_{-0.21}$ & $16.96^{+0.26}_{-0.26}$ & $16.75^{+0.28}_{-0.30}$ \\
\hline
 $a_j$ & [-5,5] & $0.60^{+0.36}_{-0.43}$ & $0.85^{+0.40}_{-0.53}$ & $-1.44^{+0.80}_{-1.08}$ & $-1.37^{+1.09}_{-1.66}$ & $-2.63^{+1.33}_{-1.21}$ & $-2.38^{+1.43}_{-1.56}$ \\
\hline
 $b_j$ & [-5,5] & $2.55^{+0.05}_{-0.05}$ & $2.52^{+0.09}_{-0.07}$ & $2.58^{+0.07}_{-0.06}$ & $2.55^{+0.05}_{-0.05}$ & $2.54^{+0.11}_{-0.09}$ & $2.54^{+0.09}_{-0.07}$ \\
\hline
 $\nu_{t,j}~[\rm MHz]$ & (0,30] & $2.82^{+0.54}_{-0.41}$ & $3.47^{+0.86}_{-0.68}$ & $0.96^{+0.21}_{-0.19}$ & $1.06^{+0.29}_{-0.27}$ & $0.91^{+0.23}_{-0.20}$ & $0.93^{+0.28}_{-0.24}$ \\
\hline
 $\Delta_j$ & (0,3] & $0.41^{+0.10}_{-0.08}$ & $0.45^{+0.16}_{-0.13}$ & $0.58^{+0.12}_{-0.10}$ & $0.49^{+0.10}_{-0.09}$ & $0.41^{+0.13}_{-0.12}$ & $0.58^{+0.14}_{-0.13}$ \\
\hline
\end{tabular}
\caption{The best-fit values (defined as the medians of the marginal distributions) and 1$\sigma$ uncertainties of the spectrum parameters for the 12 regions. 
}
\label{Table:mcmc_para}
\end{table*}

\section{Results} \label{sec:results}

In Table \ref{Table:mcmc_para} we list the fitting results of the spectrum parameters for the 12 regions. Among these parameters,  $\nu_{t,j}$ reflects the typical frequency where spectrum turnover happens. We find that regions closer to the Galactic plane have larger $\nu_{t,j}$, which means that the absorption is stronger at low Galactic latitude regions. In particular, $\nu_{t,j}$ can be up to $\sim 4.16$ MHz for regions near the Galactic Center, while in high Galactic latitude regions (i.e. $j=9,11,12$), $\nu_{t,j}$ is even lower than $\sim 1$ MHz. The 1$\sigma$ uncertainties in the derived $\nu_{t,j}$ are typically $\sim 13\%-29\%$. Thanks to the lunar occultation and the anisotropic primary beam response of the antenna, it is indeed feasible to recover the free-free absorption for different parts of the Milky Way using single-antenna observations.

In Fig. \ref{fig:nu_j} we plot $\nu_{t,j}$ of the 12 regions on the sky map. It shows more clearly that near the Galactic plane  $\nu_{t,j}$ is larger than at high Galactic latitudes. This is expected, because the density of the ISM is higher at the Galactic plane, and there are more H II regions generated from massive stars and supernova remnants (e.g. \citealt{Ne2001_1,Ne2001_2,reynolds2004warm,gaensler2008vertical,YMW16,ocker2021constraining}).
Both facts provide more free electrons, causing the stronger absorption effect.

In Fig. \ref{fig:mcmc_results} we plot the fitted spectra and uncertainties for the 12 regions, compared with the regional mean spectra of the input {\tt ULSA} sky model. We have converted the sky brightness temperature into the specific intensity using the Rayleigh-Jeans law, in order to show the absorption-induced turnover feature more intuitively. In all regions, the fitted spectrum agrees with the input sky model, and the turnover feature appears clearly. Moreover, we find that the uncertainties vary significantly among different regions and different frequencies, from  $\sim 4\%$ to $\sim 60\%$ (2$\sigma$), with the highest uncertainty appearing at the lowest frequencies.

\begin{figure}[htbp]
\centering
\includegraphics[width=0.4\textwidth]{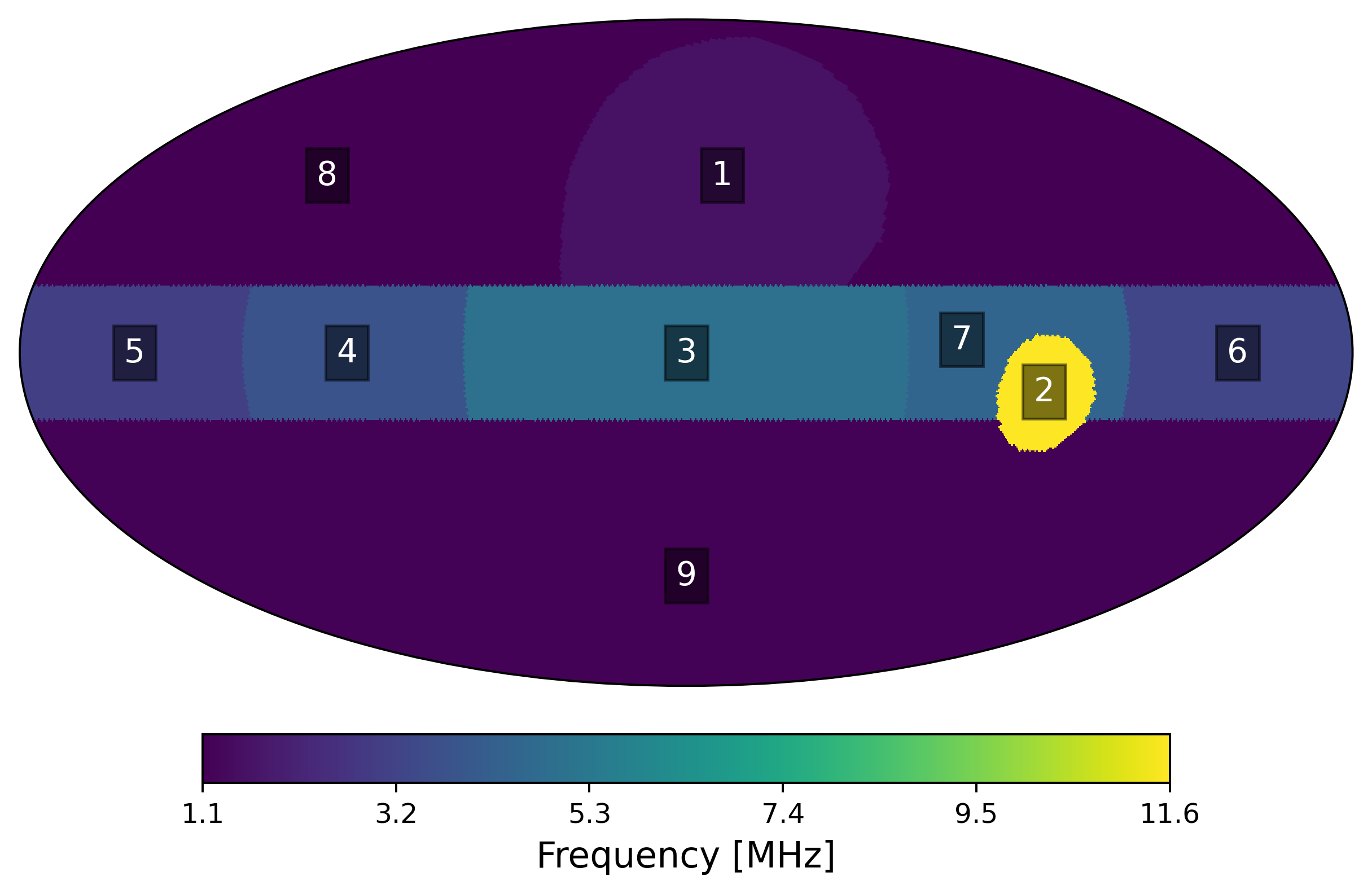}
  \caption{
  The turnover frequency $\nu_{t,j}$ for the 9 regions with labeled sequence on each pixel.
  }
  \label{fig:9 regions}
\end{figure}

\begin{figure*}[htbp]
\centering
  \includegraphics[width=0.65\textwidth]{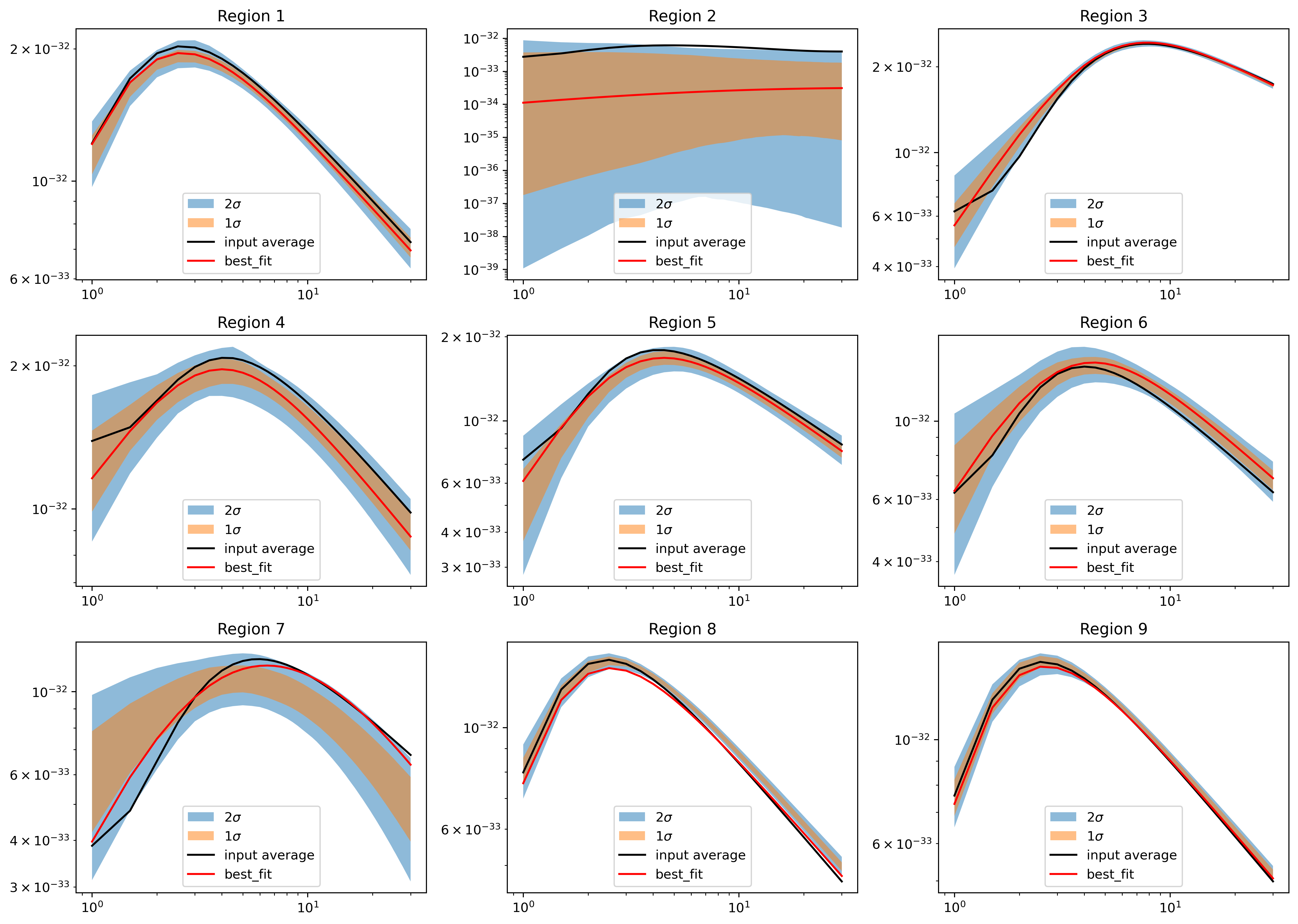}
  \caption{
  The fitted spectra for the 9 regions as illustrated in Fig.~\ref{fig:9 regions}. The line labels are the same as in Fig.~\ref{fig:mcmc_results}.
  }
  \label{fig:mcmc_results2}
\end{figure*}

To further investigate the origins of uncertainties, we manually divide the sky into 9 representative regions: Loop I, the northern high-Galactic-latitude region, the southern high-Galactic-latitude region, the Galactic center, the Gum nebula, and four other Galactic plane regions ($l\sim \pm 90^\circ$ and $l\sim \pm 135^\circ$, respectively), as illustrated in Fig. \ref{fig:9 regions}.
The best-fit values for the parameters are listed in Table \ref{Table:mcmc_para2}, and the fitted spectra and uncertainties are shown in Fig. \ref{fig:mcmc_results2}.
We find that for both Loop I and northern and southern high-Galactic-latitude regions, the best-fit spectra agree very well with the mean sky temperature, and the uncertainties are small. However, for regions close to the Galactic plane, at frequencies below $\sim 2$ MHz, Eq. (\ref{eq:fre_spectrum}) fails to describe the mean sky temperature because the absorption is strong \citep{cong_ultralong-wavelength_2021}, and this is the main source of uncertainties. For the Gum nebula, the uncertainty is obviously larger than other regions, because this exceptionally dense structure breaks our isotropic assumption made when deriving the uncertainty of Eq. (\ref{eq:C_ii3}). Moreover, it is interesting to note that Loop I is not very different from the ambient high Galactic latitude regions.

\begin{table*}[htbp]
\raggedright
\begin{tabular}{|c|c|c|c|c|c|c|}
\hline
 & prior & $j=1$ & $j=2$ & $j=3$ & $j=4$ & $j=5$\\ 
\hline
 $\log A_j~[{\log\rm K}]$ & [5,20] & $16.92^{+0.41}_{-0.53}$ & $8.81^{+3.59}_{-2.66}$ & $14.89^{+0.25}_{-0.25}$ & $15.31^{+0.49}_{-0.57}$ & $15.51^{+0.44}_{-0.50}$ \\
\hline
 $a_j$ & [-5,5] & 
 $0.43^{+0.56}_{-0.91}$ & $1.30^{+2.03}_{-3.04}$ & $0.87^{+0.22}_{-0.29}$ & $1.36^{+0.30}_{-0.38}$ & $0.70^{+0.46}_{-0.80}$ \\
\hline
 $b_j$ & [-5,5] & 
 $2.55^{+0.03}_{-0.03}$ & $2.31^{+1.69}_{-2.83}$ & $2.39^{+0.03}_{-0.03}$ & $2.56^{+0.10}_{-0.07}$ & $2.56^{+0.08}_{-0.06}$ \\
\hline
 $\nu_{t,j}~[\rm MHz]$ & (0,30] & $1.59^{+0.58}_{-0.43}$ & $11.62^{+12.26}_{-8.81}$ & $4.99^{+0.79}_{-0.73}$ & $3.78^{+1.19}_{-0.85}$ & $3.07^{+0.99}_{-0.80}$ \\
\hline
 $\Delta_j$ & (0,3] & 
 $0.47^{+0.12}_{-0.11}$ & $1.69^{+0.94}_{-1.22}$ & $0.42^{+0.08}_{-0.07}$ & $0.42^{+0.28}_{-0.18}$ & $0.45^{+0.16}_{-0.14}$ \\
\hline
 & prior & $j=6$ & $j=7$ & $j=8$ & $j=9$ &\\ 
  \hline
 $\log A_j~[\log\rm K]$ & [5,20] & $15.26^{+0.55}_{-0.59}$ & $14.42^{+0.81}_{-0.81}$ & $16.95^{+0.19}_{-0.20}$ & $16.89^{+0.15}_{-0.19}$ &\\
\hline
 $a_j$ & [-5,5] &
 $0.96^{+0.41}_{-0.68}$ & $0.61^{+0.65}_{-1.00}$ & $-0.97^{+0.82}_{-1.20}$ & $-1.03^{+0.84}_{-1.15}$ &\\
\hline
 $b_j$ & [-5,5] & 
 $2.54^{+0.11}_{-0.06}$ & $2.91^{+0.90}_{-0.31}$ & 
 $2.54^{+0.02}_{-0.02}$ & 
 $2.54^{+0.02}_{-0.01}$ & \\
\hline
 $\nu_{t,j}~[\rm MHz]$ & (0,30] & $3.29^{+1.16}_{-0.96}$ & $4.48^{+2.03}_{-1.77}$ & $1.10^{+0.28}_{-0.25}$ & $1.18^{+0.28}_{-0.26}$ &\\
\hline
 $\Delta_j$ & (0,3] & 
 $0.44^{+0.24}_{-0.18}$ & $0.85^{+1.09}_{-0.44}$ & $0.48^{+0.08}_{-0.07}$ & $0.47^{+0.06}_{-0.06}$ & \\
\hline
\end{tabular}
\caption{ 
Similar to Table \ref{Table:mcmc_para}, but the sky is divided into 9 regions with different areas as illustrated in Fig.~\ref{fig:9 regions}.
}
\label{Table:mcmc_para2}
\end{table*}

\section{Summary and Discussion} \label{sec:CC}

In this paper, we reconstruct the spectra for different sky regions from mock observational data for a single antenna onboard a lunar-orbit satellite. The satellite is assumed to be one of the DSL low-frequency satellites. Thanks to the lunar occultation of the sky and the anisotropy of the antenna primary beam response, the sky temperature received by the antenna varies with its pointing, providing us with information of sky temperature anisotropy. Moreover, orbital precession avoids the antenna repeatedly pointing to the same direction in one precession period (1.3 yr for DSL) that is much longer than the orbital period (2.3 hr for DSL). This breaks the symmetry with respect to the satellite orbital plane, providing further information about the sky temperature anisotropy. 

We generate mock observational data sample from {\tt ULSA} sky model for one precession period, and model various uncertainties, including the data averaging error, the discretization error, and the model error. We then find the best-fit parameterized frequency spectra of different sky regions simultaneously, using the MCMC method. Comparing with the input sky model, we correctly reconstructed all spectra and derived the absorption features (the turnover frequencies) for different directions of the sky. Near the Galactic plane, the turnover frequency is larger, which indicates that the absorption is stronger. At high Galactic latitude regions, the turnover frequency is smaller implying weaker absorption. Our investigation  demonstrates that a single antenna lunar-orbit experiment has the ability to reconstruct the anisotropic frequency spectrum. It will largely extend the application of the upcoming single antenna data from DSL low-frequency satellites, and any single antenna experiments either in lunar orbit or on lunar surface that work in the ultra-long wavelength band. 
This method also complements the map reconstruction method proposed in \citet{LuSEE-Night2025mapmaking}, which could also be applicable to a lunar orbiting experiment, by introducing a framework to incorporate the correlations between systematic errors in different data samples and in multiple frequencies.


In this work, we adopt a short dipole model for the antenna beam response for all frequencies from 1 MHz to 30 MHz. We have tested that a small deviation of the beam from that of a short dipole would not affect the main results. In practice, the spacecraft diffraction effect will affect the beam shape. The exact beam response in the presence of spacecraft diffraction requires careful simulations after all the details of the satellite design are fixed. This is beyond the scope of the current study. However, the exact beam shape does not affect the validity of our proposed method. It is expected that we will be able to recover the unbiased spectrum distribution of the sky as long as the beam response is known, either through simulation or measurement.

In \citet{Cong2022ApJ} it was proposed that the 3D electron distribution in Milky Way can be reconstructed from ultra-long wavelength spectra for different directions. However, that algorithm requires high angular resolution of multi-frequency sky map, up to $\sim 1^\circ$ level. The spectra reconstructed from single antenna observations cannot be directly applied in this methods. However, if the electron distribution is well parameterized (e.g. \citealt{cordes1991galactic}), the constraints on the model parameters can be derived.

\section*{Acknowledgments}
This work was supported by National Key R\&D Program of China No. 2022YFF0504300, China's Space Origins Exploration Program No. GJ11010401, and the NSFC International (Regional) Cooperation and Exchange Project No. 12361141814. 
\bibliography{main}
\bibliographystyle{aasjournal}

\end{CJK*}
\end{document}